# Identification of high-reliability regions of machine learning predictions in materials science using transparent oxide semiconductors and perovskites as examples

Evan M. Askenazi[1], Emanuel A. Lazar[2], and Ilya Grinberg[1*]

[1]Department of Chemistry, Bar-Ilan University, Ramat Gan, Israel

[2]Department of Mathematics, Bar-Ilan University, Ramat Gan, Israel

**Abstract**

Progress in the application of machine learning (ML) methods to materials design is hindered by the lack of understanding of the reliability of ML predictions, in particular for the application of ML to small data sets often found in materials science. Using ML prediction for transparent conductor oxide formation energy and band gap, dilute solute diffusion, and perovskite formation energy, band gap and lattice parameter as examples, we demonstrate that 1) analysis of ML results by construction of a convex hull in feature space that encloses accurately predicted systems can be used to identify regions in feature space for which ML predictions are highly reliable 2) analysis of the systems enclosed by the convex hull can be used to extract physical understanding and 3) materials that satisfy all well-known chemical and physical principles that make a material physically reasonable are likely to be similar and show strong relationships between the properties of interest and the standard features used in ML. We also show that similar to the composition-structure-property relationships, inclusion in the ML training data set of materials from classes with different chemical properties will not be beneficial and will slightly decrease the accuracy of ML prediction and that reliable results likely will be obtained by ML model for narrow classes of similar materials even in the case where the ML model will show large errors on the dataset consisting of several classes of materials. Our work suggests that analysis of the error distributions of ML predictions will be beneficial for the further development of the application of ML methods in material science.

**Introduction**

The application of machine learning (ML) methods to materials science and chemistry has become increasingly popular in the past decade. Both traditional machine learning methods as well as deep learning methods have been applied with some success for the prediction and design of properties of new materials, particularly in conjunction with high-throughput and combinatorial chemistry studies [1-16]. This includes optimizing feature selection and use of data decomposition methods such as MODNet [17].

Unlike physics-based methods such as density functional theory for which the limitations of accuracy can be estimated based on the approximations of the method, the black-box nature of ML methods means that it is unclear what are the limits of accuracy of any particular ML model. The fact that ML methods are essentially interpolation-based means that while accurate predictions can be obtained for many, perhaps most materials in a given class, qualitatively wrong results are likely to be obtained in some cases (Figures 1a,b). The lack of knowledge of when such qualitative failures can be expected hinders the use of ML methods for materials design. The lack of interpretability of ML results is a related problem in that a construction of an accurate ML model does not provide physical or chemical insight. This makes it difficult to extend the results of ML research to other systems and makes ML essentially another method for measuring materials properties rather than for understanding the interactions in the materials and their relationship to the desired material properties. While several recent studies have focused on addressing these problems, the development of methods for reliable and interpretable ML for materials science is still an ongoing challenge [18-19].

To increase the reliability of ML predictions, the approach of using more data and greater model complexity is often used. Inclusion of additional data is supposed to increase the sampling of the space of possible materials and provide more information to specify the parameters of the model. However, if the additional information included in the data set contains noise, it will not improve and may actually worsen the model performance. We hypothesize that since ML methods learn a complex function that depends on the features, inclusion of materials from classes with different chemical properties will have the same effect as that of noise. In other words, if materials of class A follow an analytical relationship between the chosen set of features and the property of interest, while materials of class B follow a different analytical relationship between the chosen set of features and the property of interest or have only a weak relationship between the chosen set of features and the property of interest, the inclusion of data from class B will not improve the predictive ability of the model for class A. Furthermore, if the systems in class B lack common feature-property relationships, the use of ML for systems in class B may lead to qualitatively wrong predictions that will make the trained ML model untrustworthy for class B systems even if the mean absolute error is low due to the large number of systems in class A in the dataset. This is similar to the composition-structure-property relationships that are valid for narrow classes of materials and properties and is in contrast with fundamental physical laws that are general across a wide variety of materials and compounds.

Consideration of the fact that materials datasets used for ML studies often contain a wide mix of different materials including some that are not physically or chemically reasonable leads us to hypothesize that materials datasets contain a high-reliability region in feature space such that all materials falling into this space will be predicted with high accuracy due to the similarity of their physical behavior and the relationship between the features and the target property. With increasing distance from this region, while many or even most materials can still be predicted well by a ML model, outliers will be increasingly found such that ML predictions will be much less reliable than for the materials in the high-reliability region. This also implies that for the accurate prediction of the materials in the high-reliability region, the ML model can be trained using a limited set of data in the vicinity of the region, while the inclusion of the data far away from this region in the training set will either not improve or even decrease the accuracy of the ML model. Furthermore, identification of the common features of the materials systems in the high-reliability region and the analysis of their feature-property relationships may reveal new insights into the physics and chemistry of these materials.

We propose a method for identifying the high-reliability region automatically by constructing a convex hull (CH) [20] in low-dimensional feature space that includes the maximum number of points accurately predicted by the ML model while excluding the data points with high error (Fig. 1c). Then, we can expect that the model predictions for the materials inside the convex hull will be significantly more reliable and have lower errors than the model prediction for the materials outside the convex hull, with the model prediction error increasing with increasing distance from the convex hull boundary. We examine this method for several materials datasets containing different material types and properties. Using validation sets of materials that were not used in the convex hull construction, we demonstrate that the materials that are located in the convex hull are predicted accurately while those outside the convex hull show increasingly worse prediction accuracy with increasing distance from the convex hull. We also demonstrate that for perovskite oxides where we have good physical intuition, a high-reliability region can be identified manually by simply selecting the physically reasonable compositions from the entire dataset, and that conversely convex hull construction can be used to obtain physical insight for the perovskite oxide materials.

**Results**

*Convex hull construction method*

In the search for the region of high reliability, it is necessary to balance the competing requirements of high prediction accuracy and generality. In other words, we need a method for answering the following question. What is the largest possible contiguous region in feature space for which the ML prediction error tends to decrease as one approaches the center of this region? Here, we define the prediction error of a given data point as the mean absolute error for that point obtained from a set of ML models fitted to the data set. Once such a region is defined using convex hull construction, we expect that predictions for the data in the validation data set that were not included in the convex hull construction will be highly reliable for the materials inside the convex hull and will become less reliable with the greater distance from the hull.

As a first approximation of such a region in feature space, we can take the points with sufficiently low prediction error, for example the 5% of data points with the lowest error, for the convex hull construction. However, the CH constructed using this approach include many points with high error as shown for the $ABO_3$ lattice parameter data set in Fig. S1. Therefore, a method that explicitly seeks to exclude the data points with high error from the CH and to include as many points with low error as possible in the CH is needed.

To obtain such convex hull, we define a cost function F that favors the inclusion of as many points as possible in the convex hull but imposes a penalty for including the points with errors above a certain tolerance threshold T. The cost function F is given by

$$F = -N + \Sigma_{i,N} f(e_i) \quad (1)$$

$$f(e_i) = 0 \text{ if } e_i < T \quad (2)$$

$$f(e_i) = \exp(h^* e_i) \text{ if } e_i > T \quad (3)$$

where N is the number of the points inside the convex hull, $e_i$ is the error for the $i$th data point, $T$ is the error tolerance, and $h$ is a hyperparameter that scales the rate at which the penalty for high error turns on. Thus, the optimal convex hull will include a large number of data points with low error and will tend to exclude data points with larger errors. Such an optimal convex hull can be constructed by minimizing F with respect to the included data points.

To obtain the optimal dataset, we take the initial full dataset of all systems and randomly remove a subset of data points. Then, a convex hull is drawn around the remaining points, and the quality of the convex hull is evaluated based on our designed cost function. For each iteration, the total cost function and the MAE of the points contained in the convex hull are computed; if both of them decrease relative to the MAE and cost function of the previous iteration, an additional subset of points (20 points) is removed. If only either the cost function or the MAE decrease or neither one is decreased, only 8 points are removed and 12 points that were left out are placed back in the subset used for CH construction. The process is repeated until the cost function reaches a minimum value; this provides us the optimal data set for the construction of the convex hull. The CH construction procedure is illustrated in Figs. S2 and S3. The optimal convex hull specifies the region of reliable predictions where we can expect the errors to decrease toward the center of the convex hull and to increase with the greater distance from the center of the convex hull.

We perform the ML model training and prediction and convex hull construction for material science data sets used in previous ML studies, namely the datasets of the formation energy $E_f$ and band gap $E_g$ of the $(Al_xGa_yIn_z)_2O_3$ (with $x + y + z = 1$) transition conducting oxide (TCO) compounds, dilute solute diffusion dataset, perovskite $E_f$ and $E_g$ datasets and perovskite $ABO_3$ lattice parameter dataset reported in [21-23]. Fig. 2 shows the plots of the ML-predicted versus actual values for the six datasets. The various accuracy metrics for the obtained ML models are presented in Table 1. It can be observed that for all six datasets, the ML model predicts most of the data well with most of the points falling in a narrow band around $x=y$ line, but some outliers with significantly worse accuracy are present. The prediction performance from most accurate to least accurate is in the order of TCO $E_f$, TCO $E_g$, perovskite $E_f$, dilute solute diffusion, perovskite lattice parameter, and perovskite $E_g$. It can be seen that despite the good overall performance, outliers with large errors are present for the ML model predictions for all datasets, making the ML predictions not fully trustworthy.

We then construct the convex hulls containing the high-reliability regions using the procedure described above for 80% of the data in each dataset. The points falling inside the convex hull are shown in red in Figs. 2a-e and it is observed that the data inside the convex hull form a more narrow band around *x=y* line and have fewer outliers, with the exception of the perovskite $E_g$ dataset for which the ML accuracy is noticeably worse than for the other datasets. The distinction between the full data set and the data inside the convex hull is strongest for the perovskite lattice parameter dataset and is weakest for the perovskite $E_g$ dataset. Whether a point is located inside or outside the convex hull is a crude measure of accuracy because intuitively we expect that prediction accuracy will vary smoothly with the distance from the CH boundary. Thus, we expect that the distance from the CH boundary provides a better indication of the prediction accuracy. Therefore, for the data points used in the convex hull construction, we show the ML model prediction errors as a function of distance from the boundary of the convex hull in Fig. 3 and also present error distribution plots for the data points outside the convex hull, the data points in the inner CH region (defined as at most 1/3 of the distance from the center to the CH boundary ) and the outer CH region distance (defined as located between 1/3 of the distance from the center to the CH boundary until the CH boundary) in Fig. 4.

Examination of Fig. 3 shows that for the TCO $E_f$, TCO $E_g$, perovskite $E_f$, and perovskite lattice parameter datasets, the errors are low inside the convex hull, while outside the convex hull the errors are also mostly low but a number of high-error data points are also found. The CH boundary contains systems with a large variation in the ML prediction errors. For the dilute dataset, the distinction between the inner CH region and the outside of CH is less clear, even though even for this datasets the high-error systems are also more common outside the CH than inside. For the perovskite $E_g$ dataset no distinction between the inner CH region and the outside of CH is observed. We ascribe the more poor separation of the high- and low-error systems in these datasets to the lower predictive ability of ML methods for these datasets (as shown by the wider bands around *x=y* and more outliers in Fig. 2). Such lower predictive ability suggests that the features used in the ML model are insufficient so that the property of interest is controlled by features not included in the dataset. Therefore, construction of CH based on the features used in ML model construction cannot separate the feature-space well into low- and high-error regions because the errors are controlled by additional, unincluded or hidden features.

Examination of the error distribution plots presented in Fig. 4 shows a clear separation between the distribution for the inner CH region, the outside CH region and the CH boundary region of feature space. In all cases, the inner CH region (error distribution shown in red) has a more tight distribution, and is particularly clearly distinguished from the outside the CH region (shown in green). There is no separation for the perovskite $E_g$ dataset due to the poor quality of the model predictions, as discussed above. Thus, it is clear that we can successfully isolate the region of high-reliability in 5-dimensional feature space using convex hull construction. This shows that CH construction can be used for analysis of the errors results and post-facto rationalization of why some data and materials properties are predicted better than others.

Next, we test the ability of the constructed convex hulls to describe the reliability of ML predictions for systems for which the prediction error was unavailable during CH construction. For each data set, we take the 20% of the data that were not used to construct the convex hull and calculate the distance in feature space from the convex hull boundary for these data. We then plot the error of the ML prediction for these data versus the distance from the CH boundary and error distribution plots in in Figs. 5 and 6. An examination of Fig. 5 shows that here as well, the error inside the CH is noticeably lower than the error outside with essentially all points with large errors found outside the CH boundaries for all datasets except for the perovskite $E_g$ dataset. This is also clearly observed from the error distribution plots in Fig. 6. Thus, we see that CH can be used to predict the reliability of the machine learning predictions of materials properties. The materials that fall inside the CH will have their properties predicted reliably, while for those outside the CH the reliability is more poor, with a wider variation of error values.

*Effect of extra data on the accuracy of prediction for the systems in the high-reliability region*

Next, we demonstrate that the inclusion of the data further outside the CH boundary contributes increasingly less to the accuracy of ML prediction of the data inside the CH boundary. For each data set, we systematically decrease the size of the training data by including only the data points within a certain distance from the CH boundary. We then train the ML model using this training data and then test the obtained model on the testing data *inside the CH boundary only*. To measure the impact of the added data on the accuracy of trained model, we use the change in the MAE normalized by the number of points added as defined by

$$\Delta MAE/\Delta n = (MAE_i - MAE_{i+1})/(n_{i+1} - n_i) \qquad (4)$$

where $MAE_i$ and $MAE_{i+1}$ are the MAE obtained for distances $d_i$ and $d_{i+1}$, and $n_i$ and $n_{i+1}$ are the numbers of data points with distances from the CH boundary $\leq d_i$ and $\leq d_{i+1}$, respectively. The normalization is performed because inclusion of more data points will have a larger effect on the MAE so that the change in the MAE per data point is a better metric of the impact of the data at different distances from the CH boundary. The results are shown in Fig. 7. For all data sets, the $\Delta MAE/\Delta n$ is initially very high at negative distances, indicating that insufficient data is provided for training, and then fluctuates strongly close to the CH boundary. For the data points located at normalized distances $d \geq 0.02$ past the CH boundary (see SI for distance definition), for all datasets except for the dilute solute dataset, $\Delta MAE/\Delta n$ becomes weakly negative, indicating that the inclusion of more data far from CH boundary actually decreases accuracy (albeit weakly) for the data inside the CH boundary. For the dilute solute data set, small increases in accuracy are obtained for inclusion of data at $d \geq 0.02$ as indicated by the small positive values of $\Delta MAE/\Delta n$. This indicates that for the TCO $E_f$ and $E_g$ and the perovskite $E_f$, $E_g$, and lattice parameter datasets, the materials inside the CH boundary follow feature-property relationships and physical principles or composition-property relationships that are sufficiently different from those for the materials outside the CH boundary such that the inclusion of these materials in the training data set actually decreases (albeit slightly) the prediction accuracy for the materials inside the CH boundary. For the materials in the dilute dataset, the materials outside the CH boundary are still somewhat similar in their feature-property relationships and physical principles or composition-property relationships to the materials inside the CH boundary. Thus, analysis of prediction accuracy changes with the changes in the training set can identify groups of materials that are physically and chemically similar to each other.

*Using physical understanding to identify high-reliability regions*

Next we examine the connection between the high-reliability ML prediction regions found by our procedures and the physical and chemical principles that govern the properties of materials. Here, we focus on the perovskite oxide lattice parameter dataset because we have good physical intuition for $ABO_3$ perovskites derived from the long-standing research effort devoted to understanding these materials. In particular, the lattice parameter is one of the most basic materials properties and good physical intuition has been obtained during the investigations of $ABO_3$ systems starting with the pioneering work of Goldschmidt in the 1920s [24] facilitating the understanding and interpretation of ML results. For the lattice parameter prediction, we use the lattice parameter values obtained for high-symmetry 5-atom ABO3 perovskites by density functional theory calculations for all possible combinations of A and B elements, as reported in a previous work [25].

As described in the introduction, intuitively the materials that follow the same set of physical and chemical bonding principles should exhibit similar behavior and be predicted accurately by ML models. Therefore, for well-understood properties such as the lattice parameters of $ABO_3$ oxides, it should be possible to identify high-reliability regions manually based on the understanding of the relationship between the features in the dataset and the target property (ABO3 lattice parameter in our case).

An inspection of the plot of the ML-predicted versus actual lattice parameter values shows that the majority of points fall within an error band of approximately ±0.1 Å (Fig. 8a). Similarly, a plot of the frequency distribution of the errors shows that a Gaussian distribution with a full width at half-maximum (FWHM) of 0.05 Å (Fig. 8b), providing another estimate of the uncertainty of the ML prediction. Furthermore, several outliers are observed with errors of 0.1-1.0 Å, with the maximum error of 1.05 Å (Fig. 8a). For the majority of the $ABO_3$ systems, the ML predicted values are highly accurate with an MAE of 0.026 Å corresponding to ~0.5% of the lattice parameter, which is comparable to the lattice parameter errors of generalized gradient approximation (GGA) functionals. Nevertheless, for about 100 systems, lattice parameter errors are 0.1-0.2 Å, corresponding to errors of 2.5-5%, which is quite high. In addition, 35 qualitative failures are obtained with lattice errors of ≥7%, corresponding to cubic volume error of 21%.

We use physical principles to identify the high-reliability region for the $ABO_3$ lattice parameter data set. Empirically, cubic perovskite $ABO_3$ systems that can be synthesized at standard pressure have been found to obey two criteria. First, the perovskite must be charge-balanced, such that the formal charges of the A and B sites are equal to 6, and second it should exhibit a good match between the preferred A-O and B-O bond distances in the perovskite structure, as characterized by the Goldschmidt tolerance factor t given by

$$t=(R_A+R_O)/((R_B+R_O)\sqrt{2}) \qquad (5)$$

Here, $t<1$ ($>1$) indicates that the A-O perovskite sublattice prefers smaller (larger) lattice parameter than the B-O sublattice, while $t=1$ indicates a perfect match between the preferred lattice parameters of the two sublattices. Typically, perovskites that can be synthesized at ambient pressure have $t$ values between 0.9 and 1.1. Thus, intuitively the physically realistic charge-balanced ABO$_3$ materials with 0.9<$t$<1.1 are likely to follow different composition-property relationships than the ABO3 materials that lack charge balance and have tolerance factors outside the 0.9<$t$<1.1 region. Furthermore, because these materials follow well-known physical and chemical principles, we expect that they will be predicted accurately, whereas for the non-charge-balanced ABO$_3$ or ABO$_3$ with either too large or too small tolerance factors, prediction accuracy will be more poor due to the wide range of possible behaviors in these systems.

To examine this hypothesis, we compare the lattice parameter errors for the full ABO$_3$ data set, the charge-balanced subset of the ABO3 data, and the physically realistic (charge-balanced and the 0.9< $t$ <1.1) subset of the ABO$_3$ data (Fig. 8a). It is clear that the charge-balanced subset shows lower error than the full ABO$_3$ data set, and the physically realistic subset shows lower errors than the charge-balanced subset. The MAE and standard deviation of the errors are reduced from 0.026 Å and 0.0469 Å for the full ABO$_3$ data set to 0.021 Å and 0.03 Å for the charge-balanced subset and to 0.018 Å and 0.0187 Å for the physically realistic subset. In particular, no extreme outliers with errors above 0.5 Å are present in the charge-balanced subset and the maximum error for the realistic subset is only 0.11 Å. Thus, it is clear that the realistic subset data form a high-reliability region.

The degradation of the accuracy in the ML model prediction away from the more physically reasonable region in feature space can also be observed in the plot of the MAE, FWHM and MaxAE ML model errors as a function of the sum of cation charges and as a function of the tolerance factor for the charge balanced systems (Figures 8c,d). It can be seen that the errors increase with greater difference of the charge sum from 6 and with the greater deviation of the tolerance factor range from 0.9<$t$<1.1. These results validate our hypothesis that the model is learning a physically based relationship that is more accurate for the polar covalent cation-O bonding in the physically reasonable charge-balanced and lattice-matched ABO$_3$ systems.

Similar to the results observed in Fig. 7, the accuracy of ML prediction for the systems in the high-reliability region (realistic subset ) is not improved by including ABO$_3$ systems far away from the boundary of the high-reliability region. Fig. 8e shows the results for the $\Delta$MAE/$\Delta n$ for models trained on the ABO$_3$ lattice parameter data sets created using tolerance factor and charge balance as cutoff criteria for inclusion in the training data. As shown in Figure 8d, as the range of the $t$ values included in the training set increases, $\Delta$MAE/$\Delta n$ values drop steeply between 0.875 Å< $t$ < 1.125 Å and 0.85 Å< $t$ < 1.15 Å, and then become slightly negative when training on the full set. Then, the prediction accuracy for realistic ABO$_3$ systems decreases slightly as more unrealistic systems are added to the dataset. Thus, inclusion of these additional ABO$_3$ systems does not help and in fact slightly hinders the fitting of an accurate ML lattice parameter model for the high-reliability region in the feature space that corresponds to realistic ABO$_3$ systems.

*Extracting physical understanding from CH hull identification*

Our finding that well-known chemical and physical principles can be used to identify the high-reliability region of ML prediction in feature space raises the question of whether conversely, the identification of high-reliability region of ML prediction in feature space can be used to reveal as-yet-unknown common chemical and physical principles or similarities for the materials identified to belong to the high-reliability region.

Analysis of the CH found by the minimization of the cost function given by Eq. 1 shows that this CH encloses the systems that are mostly charged balanced with some systems with small deviations from the cation charge sum of 6. However, these systems have low tolerance factor values between 0.8 and 1.0 (Fig. S4). (We also found a CH that mostly overlaps with the experimentally viable high-reliability region identified manually, as described in the SI.}

To understand the difference in the interactions between the A, B and O atoms that give rise to the preferred $ABO_3$ volume between the sets of $ABO_3$ systems enclosed by the CH and identified based on physical arguments, we fit the volumes of the $ABO_3$ systems for the two data sets separately to the formula proposed by Sidey [26]  We obtain

$$a = 0.31 R_A + 1.63 R_B + 1.79 R_O \quad (6)$$

and

$$a = 0.71 R_A + 1.66 R_B + 1.40 R_O \quad (7)$$

for the higher $t$ ($0.9 \leq t \leq 1.1$) $ABO_3$ high-reliability region and the lower $t$ ($0.8 \leq t \leq 1.0$) $ABO_3$ high-reliability region, respectively. It is observed that the coefficients describing the effect of the B-site and O atoms on the $ABO_3$ volume are similar for both datasets, while the coefficient for the A-site for the CH set is greater by almost a factor of two than that for the  higher $t$ set.  In other words, for the systems in the higher $t$ data set identified by physical intuition, the lattice volume is relatively insensitive to the size of the A-site atom while for the lower $t$ set, the lattice volume depends more strongly on the A-site ion size.  For $t < 0$, the lattice volume is determined by the competition between the preferences of the B-site cation for longer bonds in a larger lattice and the preference of the A-site for shorter bonds in  smaller lattice. Therefore, the higher coefficient for $R_A$ in Eq. 6 compared to that in Eq. 5 means for the systems in the CH set, the A-O bonds are more stiff relative to the B-O bonds than for the systems in the experimentally viable set. Since a smaller $t$ indicates a bigger mismatch between the small A-site and large B-site, for small A-site cations and large B-site cations the A-O bonds will be relatively more stiff compared to the B-O bonds than for larger A-site cations and smaller B-site cations.  This is consistent with the  well-known concept of chemical hardness that describes the stiffness of the bond and which tends to increase with smaller size of the element. This suggests that the lattice volume of cubic $ABO_3$ oxides depends on the chemical hardness of the elements due to its important for bond stiffness. Thus, the identification of the two different regions of high ML reliability reveals the difference between the interatomic interactions and composition-structure relationships governing the $ABO_3$ oxides in these regions.

**Discussion**

Our findings have the following implications for the broad field of ML in materials science.

First, our results show that regions in feature space with high prediction reliability can be identified; for these sets of systems the ML model predictions are reliable because these systems follow similar feature-property relationships. The CH method can find the high-dimensional boundary that encloses such a high-reliability region in feature space.

Second, inclusion of data outside the high-reliability region in ML model training does not increase the accuracy of prediction for the systems inside the high-reliability region due to the different feature-property relationships followed by the systems inside the high-reliability region and systems outside this region. Similarly, the traditional MAE, MSE and MaxAE metrics computed for the full data set underestimate the accuracy of ML models in the region of reliability for which the ML methods are suited.

Third, a high prediction error for a system that is located inside the high-reliability region in feature space suggests that the properties of this systems are governed by features not included in CH construction or in the training of the ML model. Thus, CH construction can be used to identify systems for which behavior is controlled by "hidden" features omitted from the ML model construction.

Fourth, the ML model are likely to be more accurate for physically relevant systems that obey usual chemical and physical rules. This is in contrast to the systems that do not obey the usual chemical and physical rules that due to a much greater possible variability of feature-property relationships are less likely to be predicted accurately. This is an example of the Anna Karenina principle previously applied in biology and social science [27-29].

Fifth, analysis of ML results can be used to identify chemically similar regions and extract physical understanding. This means that ML can be used as method for scientific examination of classes of materials to elucidate physical and chemical relationships rather than as a purely predictive data modeling tool.

Finally, our work suggests that to improve the usefulness of ML methods in materials science, in addition to improving prediction accuracy, it will be beneficial to focus on the development of methods for the analysis of the reliability of ML results.

**Conclusion**

We have investigated the reliability of ML predictions for several materials science datasets. We find that a convex hull that includes the maximum number of well-predicted data while excluding the poorly-predicted data can be constructed to identify the high-reliability regions in feature space for these datasets. The constructed convex hulls enclose systems that are physically and chemically similar and therefore have similar feature-property relationships that can be predicted well by ML models. We demonstrate that in addition to the identification of similar systems, the obtained convex hulls well predict the accuracy of ML prediction for systems that were not included in the convex hull construction, with lower (greater) prediction accuracy obtained with more positive (negative) distance from the CH boundary. Thus, CH construction can be used for estimating the reliability of ML predictions. Analysis of the ML results shows that the ML predictions are more reliable for more physically relevant systems, and that the ML prediction accuracy for the systems in the CH region is not increased and rather is actually slightly worsened by inclusions of data of the systems far from the CH boundary. Our work suggests that analysis of the error distributions of ML methods will be beneficial for the further development of the application of ML methods in material science.

**Methodology**

*Datasets and ML Model*

In this work, we focus on six datasets used in previous ML materials science studies, namely the datasets used for the prediction of transparent semiconductor formation energy $E_f$ and band gap $E_g$ [21], the dataset use for the prediction of dilute solute diffusion [22], datasets used for the prediction of perovskite oxide $E_f$ and $E_g$ [23] and the dataset of lattice parameter of all possible perovskite oxides. For each dataset, we use the XGboost gradient boosting method [30-32] and the support vector regression (SVR) to construct ML models for predicting the properties, with the results for the more accurate method (XGboost for the $ABO_3$ lattice parameter, and SVR for the other datasets) used in this study. The hyperparameters of each method are optimized with 10-fold cross-validation. For all data sets used in model fitting, the data is split such that 90% are used for training and the remaining 10% form the test set. For each subset, an ML model is trained, with the aforementioned hyperparameter optimization through 10 fold cross validation, on the other 9 subsets and tested on the selected subset. A total of ten models are thus trained, one for each subset. In this manner, prediction for all of the materials in the datasets are made. The features used for the transparent semiconductor formation energy $E_f$ and band gap $E_g$ [21], the dataset used for the prediction of dilute solute diffusion [22], and datasets used for the prediction of perovskite oxide $E_f$ and $E_g$ [23] are the same as in previous work. For perovskite oxide lattice parameter predictions, we construct a highly accurate ML model through the choice of representative features. We use element labels, ionic radii and valence charge values, electronegativity and the periodic table block of the A and B site ions as the regression features. Since ionic size depend on the valence of the ion and its coordination, we use ionic radii for the two most common ionic valences in the 12-fold coordination for the A-site and in the 6-fold coordination for the B-site. If only one ionic valence/charge value is possible, only that value is used. The valence charge numbers and the periodic table block of the ions (s,p,d) are used as numerical label classifiers. For the prediction of the perovskite oxide lattice parameters, the ionic sizes, charges, the nature of the valence orbitals (s,p,d,f) and were used as features. The developed ML model for perovskite oxide lattice parameters is described in detail and compared to previously reported models in the literature in the SI. The accuracy of the obtained ML models for all six datasets is illustrated in Fig. 2 where the predicted values are plotted versus the actual values.

*Convex hull construction*

For a given set of data points that are functions of n features, the *convex hull* is defined as the smallest region in the n-dimensional feature space that includes the data points and all of their weighted averages, as illustrated for two dimensions in Figure 1c. Methods for automatically finding this region have been developed and are publicly available [33]. Currently, the highest dimensionality of the available convex hull construction method is five and therefore we use these methods in the feature space formed by the five most important features (or the first five principal components) for each dataset to obtain a convex hull that includes as many as possible data points with low prediction error. For each dataset, only 80% of the data are used for convex hull construction, with remaining 20% of the data held for validation.

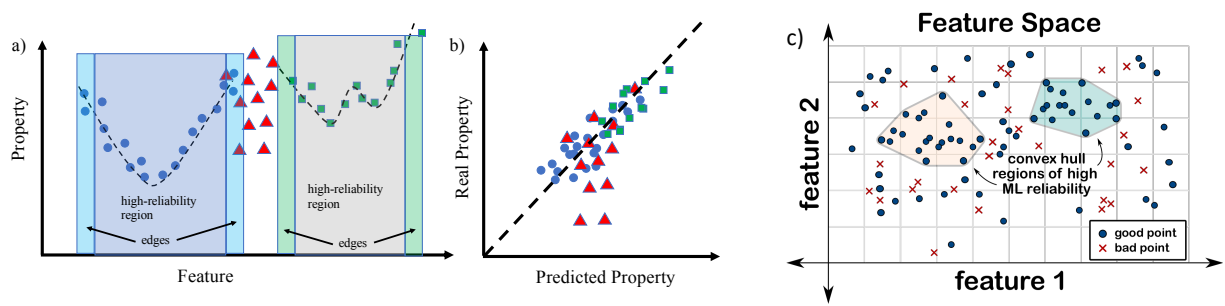

**Figure 1.** a) Schematic illustration of high-reliability regions for property prediction through machine learning methods. The high-reliability regions together with their edges that are necessary for accurate predictions are marked by rectangles. b) Real versus ML-predicted property for a case where high-reliability regions exist together with low-reliability regions. Data points from the two high-reliability regions are shown in blue and green. Data points from the low-reliability region are shown in red. c) Schematic illustration of the convex hull regions enclosing the greatest set of high-quality predictions in feature space while excluding low-quality predictions.

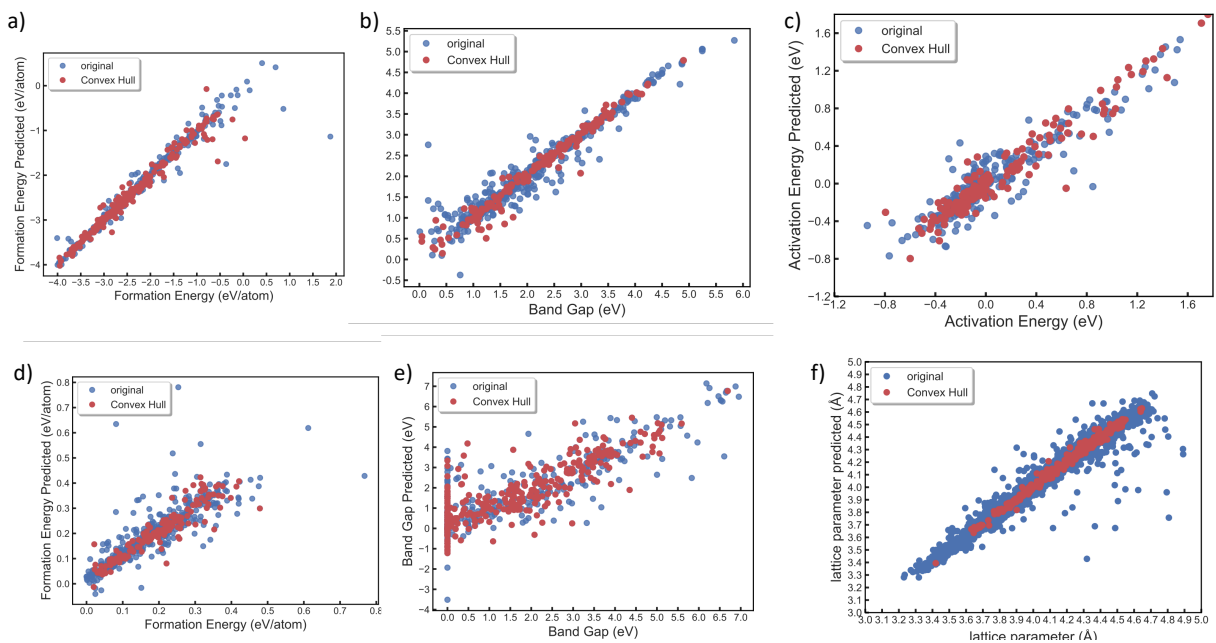

**Figure 2. Prediction accuracy of ML models** a) TCO $E_f$ dataset b) TCO $E_g$ dataset c) Dilute solute diffusion dataset d) Perovskite $E_f$ dataset e) Perovskite $E_g$ dataset f) Perovskite oxide lattice parameter dataset. ML predictions for the full datasets are shown in blue, ML predictions for the systems included in the constructed convex hulls are shown in red.

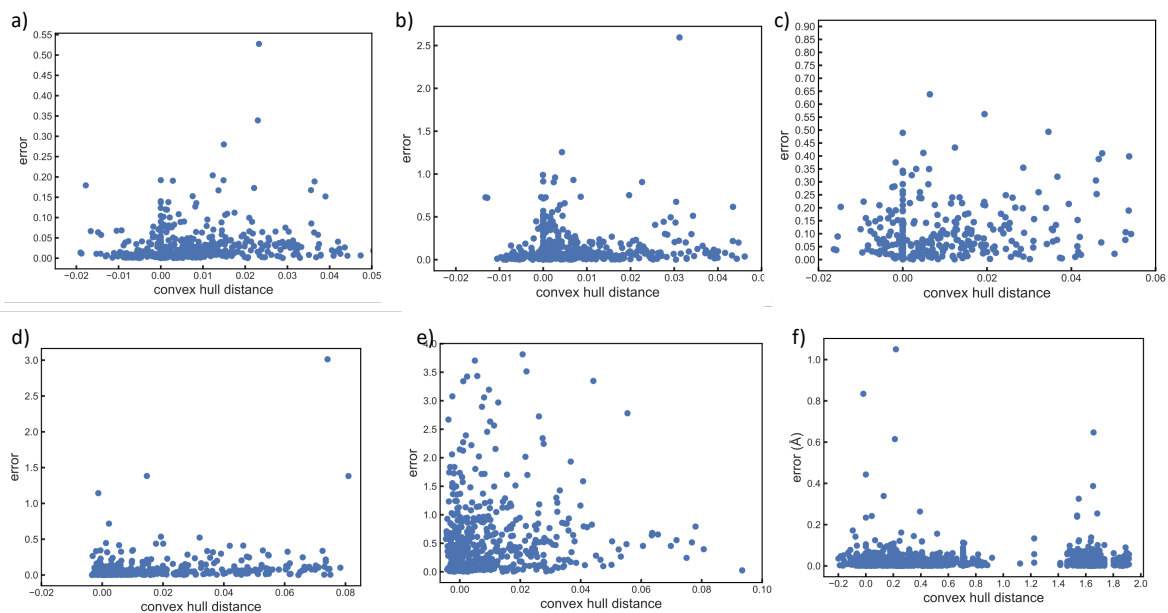

**Figure 3. Prediction error of ML models as a function of distance from convex hull boundary for the systems used in the convex hull construction** a) TCO $E_f$ dataset b) TCO $E_g$ dataset c) Dilute solute diffusion dataset d) Perovskite $E_f$ dataset e) Perovskite $E_g$ dataset f) Perovskite oxide lattice parameter dataset.

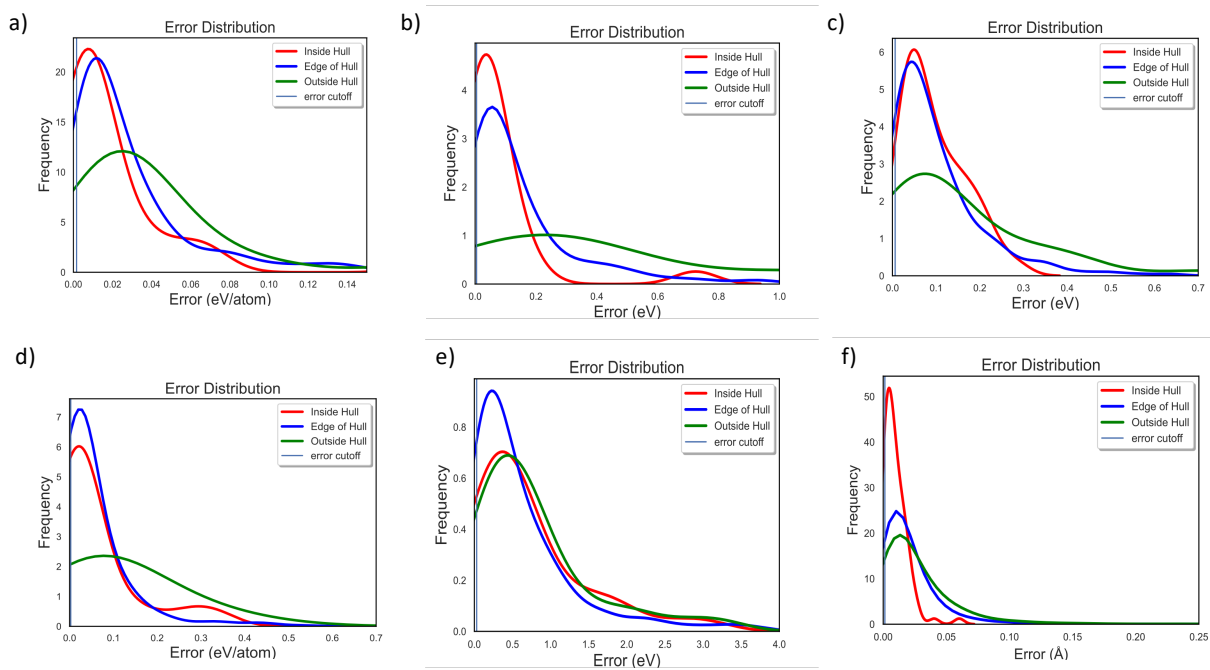

**Figure 4. Error distributions of ML models for different regions of feature space relative to the hull boundary for the systems used in the convex hull construction** a) TCO $E_f$ dataset b) TCO $E_g$ dataset c) Dilute solute diffusion dataset d) Perovskite $E_f$ dataset e) Perovskite $E_g$ dataset f) Perovskite oxide lattice parameter dataset.

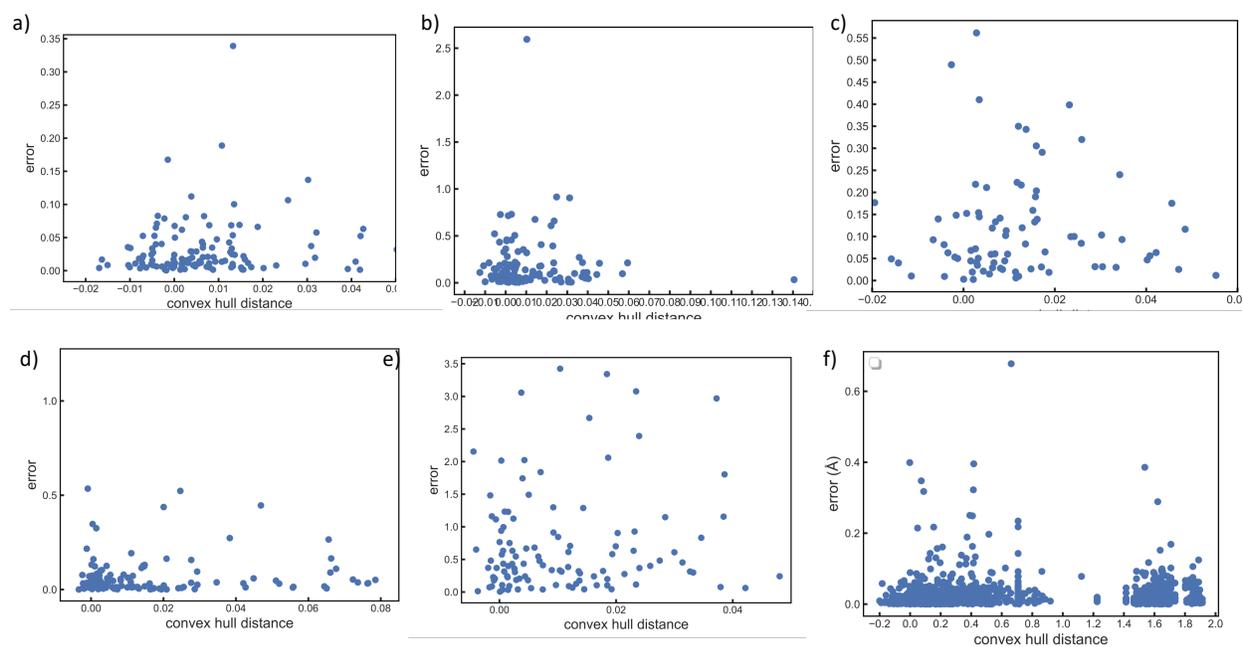

**Figure 5. Prediction error of ML models as a function of distance from convex hull boundary for the systems not used in the convex hull construction** a) TCO $E_f$ dataset b) TCO $E_g$ dataset c) Dilute solute diffusion dataset d) Perovskite $E_f$ dataset e) Perovskite $E_g$ dataset f) Perovskite oxide lattice parameter dataset.

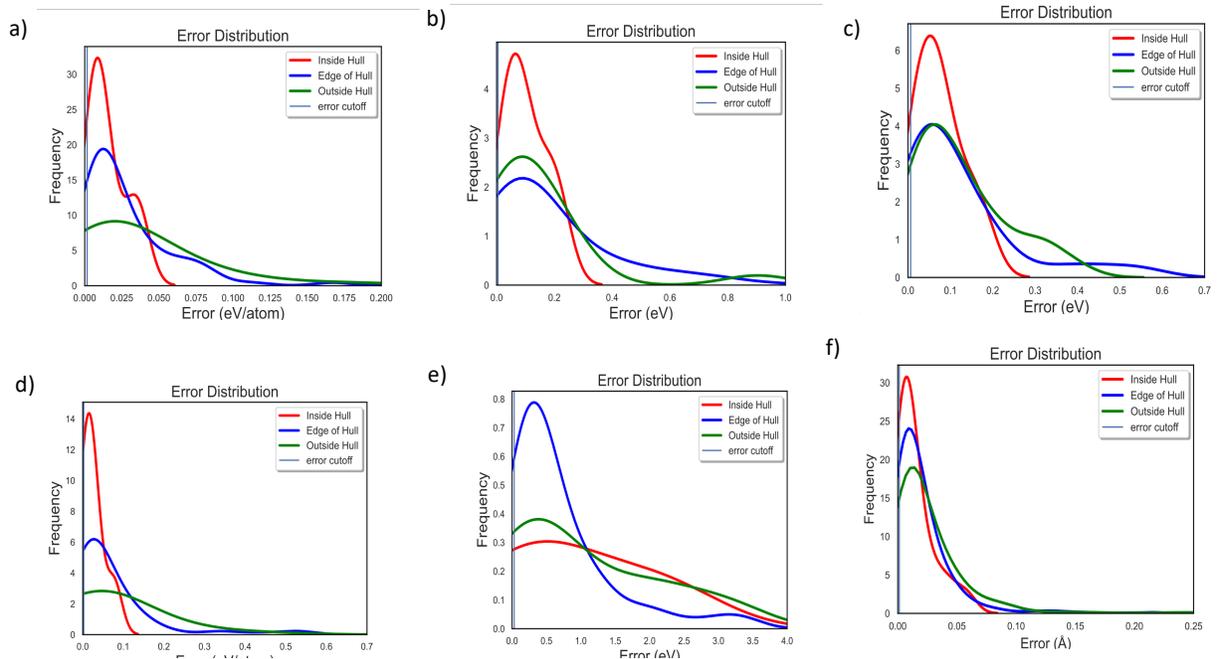

**Figure 6. Error distributions of ML models for different regions of feature space relative to the hull boundary for the systems not used in the convex hull construction** a) TCO $E_f$ dataset b) TCO $E_g$ dataset c) Dilute solute diffusion dataset d) Perovskite $E_f$ dataset e) Perovskite $E_g$ dataset f) Perovskite oxide lattice parameter dataset.

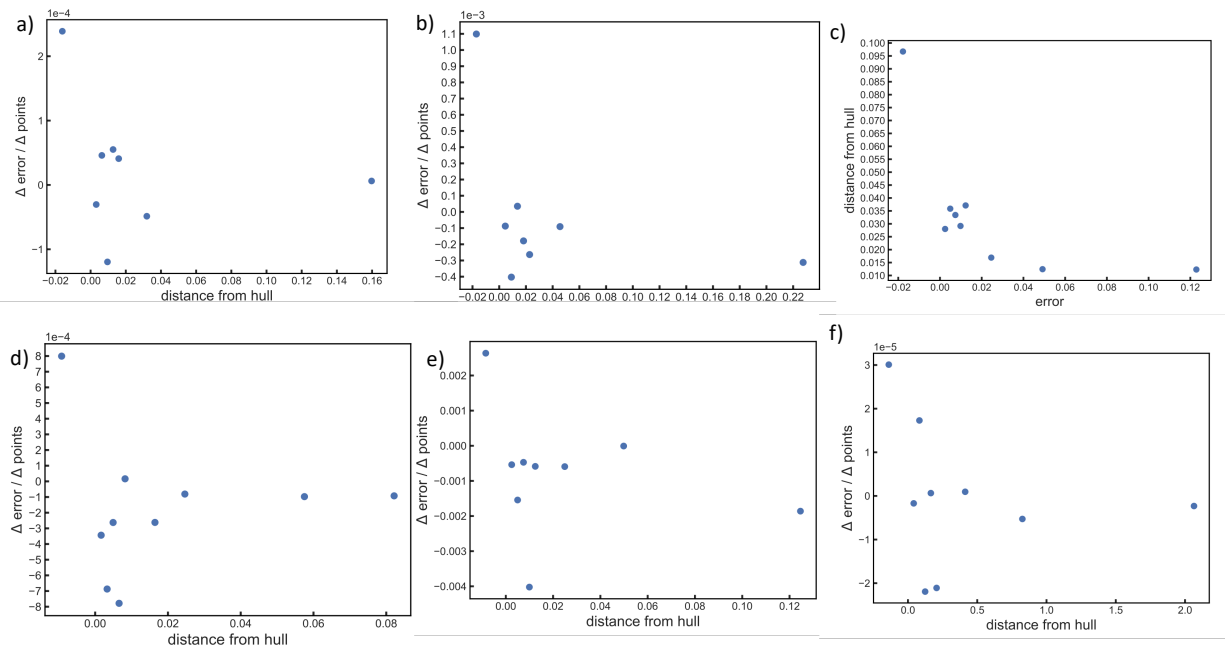

**Figure 7. Normalized change in the ML model prediction accuracy for data inside the convex hull with inclusion of data points at different distances from the hull boundary in the training set** a) TCO $E_f$ dataset b) TCO $E_g$ dataset c) Dilute solute diffusion dataset d) Perovskite $E_g$ dataset e) Perovskite $E_g$ dataset f) Perovskite oxide lattice parameter dataset. The data are sorted by distance from the CH boundary and then are split into 8-10 groups with equal number of points in each group.

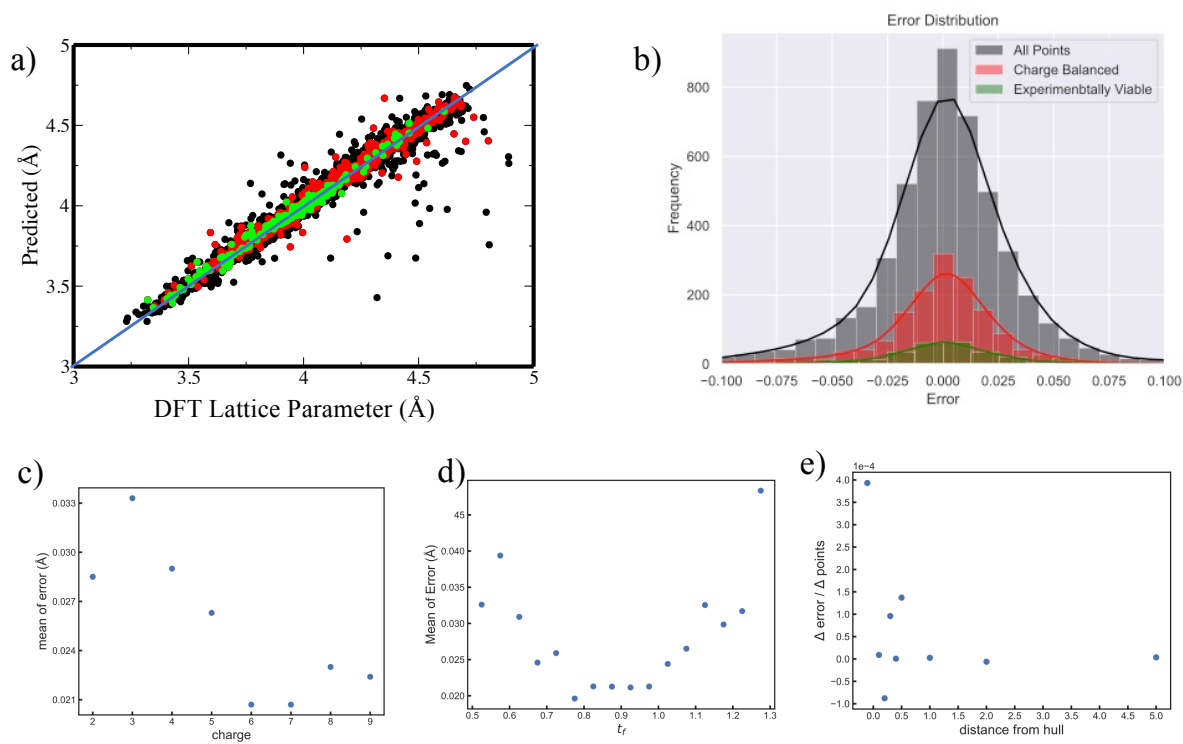

**Figure 8. Identification of high-reliability region from physical principles for ABO$_3$ perovskite lattice parameter prediction** a) Predicted versus DFT lattice parameters for our ML model trained on all ABO$_3$ systems and tested for all ABO$_3$ system (black), charge-balanced systems (red), and experimentally viable systems (green). b) Error distributions for all ABO$_3$ system (black), charge-balanced systems (red), and experimentally viable systems (green). c) MAE of the ML model predictions for ABO$_3$ system as a function of total A and B formal charge. d) MAE of the ML model predictions for ABO$_3$ system as a function of tolerance factor. e) Normalized change in the ML model prediction accuracy for data inside the convex hull with inclusion of data points at tolerance factor ranges distances

## Data Availability

The raw and processed data that support the findings of this study are available from **E.M.A.** upon reasonable request.

## Code Availability

The code used to generate the findings of this study is available from **E.M.A.** upon reasonable request.

**Competing Interests**

The Authors declare no competing financial or non-Financial Interests

**Author Contributions**

**E.M.A.** performed the computational work. **E.M.A**. and **I.G.** jointly designed the methods and analysed the calculations. **E.A.L**. introduced the convex hull algorithm for finding optimal subspaces for predictions of materials' properties. All authors contributed to writing the paper and making the figures.

Supplementary information for

# Identification of high-reliability regions of machine learning predictions in materials science using transparent conducting oxides and perovskites as examples

Evan M. Askenazi[1], Emanuel A. Lazar[2], and Ilya Grinberg[1*]

[1]Department of Chemistry, Bar-Ilan University, Ramat Gan, Israel

[2]Department of Mathematics, Bar-Ilan University, Ramat Gan, Israel

*S1. Convex hull construction*

As a first approximation of for identifying a high-reliability region in feature space, we take the points with sufficiently low prediction error, for example the 5% of data points with the lowest error, for the convex hull construction. However, as shown for the $ABO_3$ lattice parameter data set in Fig. S1, this approach does not provide good separation between the high-reliability region containing only well-predicted systems and the rest of the feature space that contains both well- and poorly-predicted systems. As can be seen in Fig. S1a,b the error distributions for the systems inside the constructed CH and for the systems outside the constructed CH are the same for both the training set (80% of the systems included in CH construction) and the test set (20% of the systems not included in the CH construction. Thus, no enhancement in reliability is obtained inside the CH. Similarly, the error distributions as a function of the distance from the CH boundaries for the training and test set show that high prediction errors are often obtained inside the CH. Thus, it is clear that a method that constructs the CH while explicitly excluding the poorly predicted systems is necessary.

Therefore, we introduce a cost function for including systems with higher error and choose the points for CH construction such that the cost function is minimized as described by Eqs. 1-3 in the main text. The flow chart for our CH construction procedure is shown in Fig. S2 and the minimization of the introduced cost function and the MAE for the lattice parameter data set during the CH construction procedure as a function of iteration is shown in Fig. S3. It can be seen that the optimal convex hull has the cost function below 0 and the MAE is approximately 0.014 Å.

*S3. Analysis of charge balance and tolerance factors of systems in the high-reliability CH region*

To identify the common properties of the systems in the perovskite lattice parameter dataset identified to be located in the high-reliability region in feature space, we plot the total charge and tolerance factor distributions of the systems inside and outside the CH boundary in Fig. S4a,b. It can be seen from figure S4a that the systems inside the CH boundary are close to charge balanced (with total cation charge of 5or 6) with only a few systems with larger (±2) deviations from charge balance. By contrast, the systems outside the CH boundary show a much broader distribution of total cation charge. Fig. S4b shows that most of the systems inside the CH boundary have tolerance factors in the range of 0.7-0.95, while the systems outside the CH boundary have a wide range of tolerance factors.

*S4. Analytical models and previous ML models for prediction of $ABO_3$ lattice parameters*

Prediction of perovskite lattice parameters has been studied both using analytical [26,34-35] and statistical and artificial intelligence based methods [36-42]. As shown in Fig. S5, the analytical methods capture the rough trend of the lattice parameter changes between different systems but show large errors with MAE of approximately 0.14 Å. Artificial neural networks (ANN), Support Vector Machines (SVR), Convolutional Neural Networks (CNN) and General Regression Neural Networks have been the primary techniques for this analysis. The Crystal Graph Convolutional Neural Network (CGCNN) has also been used. In the CGCNN method, a graph network consisting of nodes containing ionic information is used to predict various structure properties. Hirshfeld surfaces, due to their use in the analysis of the structure of atoms bound in molecules[43-46] have been used in conjunction with AI methods as well. CNNs have been combined with these surfaces [42] which provide molecular shapes for compounds by summing the average electron densities of the atoms within the compound and within a reasonable short distance away from it. To complement the Hirshfeld surface, the contact distances from each point on the Hirshfeld surface to the closest atoms inside and outside the surface can be converted to a set of two dimensional fingerprint plots. This leads to a set of input data on which the CNN can then be trained.

A survey of the performance of these methods shows that the mean absolute error (MAE) of ML methods is between 0.026 and 0.04 Å while the MAE of analytical methods is approximately 0.14 Å.

*S5. Our XGboost model for prediction of $ABO_3$ lattice parameters*

To obtain a high-quality ML model, in this study, we use an ensemble-based XGboost machine learning method trained on a set of input features based on the key properties of the individual A and B site ions of cubic oxides. We use element labels, ionic radii and valence charge values, electronegativity and the periodic table block of the A and B site ions as the regression features. Since ionic size depend on the valence of the ion and its coordination, we use ionic radii for the two most common ionic valences in the 12-fold coordination for the A-site and in the 6-fold coordination for the B-site. If only one ionic valence/charge value is possible, only that value is used. The valence charge numbers and the periodic table block of the ions (s,p,d) are used as numerical label classifiers. The set of oxides used for model training is obtained from OQMD [34] and includes 5250 $ABO_3$ systems. For all data sets used in model fitting, the data is split such that 90% are used for training and the remaining 10% form the test set. Additionally, for the 5250 oxides, the testing data sets are randomly split into 10 subsets such that each oxide is in only one of the ten subsets. For each subset, an Xgboost model is trained, with the aforementioned hyperparameter optimization through 10 fold cross validation, on the other 9 subsets and tested on the selected subset. A total of ten models are thus trained, one for each subset. In this manner, a lattice parameter prediction for each of the 5250 oxides is made.

The ability of the Xgboost model to predict the lattice parameters for oxides under various conditions is shown in Fig. S5a. As shown in Table 1, when trained on the full $ABO_3$ data set and used for experimentally viable oxides, the model outperforms the analytical formulas and obtains a substantially smaller MAE than a recent support vector based model and deep learning model, despite using a small set of features. For mean square error (MSE), an order of magnitude reduction in the error is obtained. Displaying the residuals, as shown in Fig. S5b for a test set of computational oxides, illustrates the models' effectiveness in fitting for experimentally feasible and unfeasible oxides. Its MAE is 0.025 Å and its MSE is 0.0015 Å$^2$. Its overall error is lower than previous attempts at AI based lattice parameter predictions, albeit with greater likelihood of outliers. More than 95% of the lattice parameter predictions have an error less than 0.1 Å.

The feature importance and plot of error relative to number of features are presented in Figs. S5c and S5d, and the list of the most important features for each number of features is presented in Table 2. Feature importance was computed using Gini selection while the optimal features for each specific number of features were determine using backwards recursive feature selection. The greater importance of the B site properties relative to those of the A site is evident from the feature importance plot when all features are utilized in the selection process. This may be due to the B site features having substantially greater variance than the A site features. Alternatively, this may be due because the B-O bonds are stronger than A-O bonds and therefore are more important for determining the $ABO_3$ lattice parameter. However, for optimizing the feature set for a given number of features, A and B site properties are both necessary.

The first major drop in the error is observed when two features are used. In this case, the optimal feature set consists of the B site label and the smaller value for the A-site ionic radius. With three features, the B-site label, the A site electronegativity, and the smaller value of the two values of the A-site ionic radius are included. When the desired number of features is set to five, the second major drop in the error is recorded, and the higher of the two valence charges of the A site is added as a feature. Therefore, the usefulness of features in both the A site and B site is evident in terms of obtaining the optimal fit. It is clear that no more than nine features are necessary to reach the limit of ML model accuracy. The excellent performance of our method compared to the previous ML studies, including the deep learning model based on the Hirshfield surface indicates that our feature set is well-suited for modeling the lattice parameters of perovskite oxides, and provides another example of the crucial role of feature selection in the application of ML to small materials science data sets.

*Comparison of the effect of the choice of training data for the prediction of the systems in the high-reliability region of $ABO_3$ lattice parameter data set feature space*

We now investigate whether the $ABO_3$ systems lying outside this region of reliability contribute to the prediction accuracy of the data points in the reliability region of physically reasonable $ABO_3$ oxides. Intuitively, based on the illustration shown in Figure 1, we expect that some systems outside the limits of the region of reliability must be included for the accurate interpolation of the $ABO_3$ systems close to the edge of the region of reliability. First, in Table S3 we present the MAE, stdError and MaxAE metrics for the full data set, the subset of perovskite oxide systems in the region in feature space enclosed by CH, the manually selected subset of systems chosen based on meeting the experimental viability criteria (0.9<t<1.1 and charge-balance) and the joint subset containing both the CH and experimentally viable system. In all cases the XGboost model was trained on the full data set. A reduction in all error metrics is observed going from the full data set to the high-reliability subsets.

Then, we also train three Xgboost models using only the systems in the CH subset, the experimentally subset or the joint subset, respectively, and then test for the accuracy of model prediction for each of the full set and the three data set, thus obtaining a 4x4 matrix of MAE results for lattice parameter prediction as shown in Table S4. Examination of the MAE data presented in Table S4 shows the following. First, the high-reliability regions identified by the CH procedure and manually contain systems that follow different relationships between the features and the lattice parameter. This can be seen from the very high MAE obtained when training on the systems in the CH region and testing on the experimentally viable systems (0.096 Å) or vice versa (0.088 Å). Second, when the model is trained on systems from both high-reliability regions a fairly low MAE of 0.017 Å is obtained showing that when given the information about both subsets, the Xgboost model can distinguish between the two different regions in feature space and provide appropriate predictions for both subsets. As expected, training only on the systems in the high-reliability regions leads to poor prediction ability for the rest of the dataset as shown by the high MAE value of 0.061 Å obtained when training on both CH and experimentally viable systems and testing on all oxides. Third, confirming the conclusions from the trends presented in Fig. 7f and 8c-d, we find that some systems outside the high-reliability region must be included in the training for maximum accuracy. Therefore, training using the full set leads to better prediction accuracy for the experimentally viable high-reliability region (MAE of 0.018 Å) than when only the data in the experimentally viable high-reliability region are used for training (MAE of 0.022 Å). Finally, the high-reliability region identified by the CH construction procedure gives slightly smaller MAE than the manually identified high-reliability region, with MAE values of 0.018 Å and 0.015 Å, respectively, obtained when using the full oxides data set for training, and MAE values of 0.014 Å and 0.022 Å, respectively, when using both the CH and experimentally viable data subsets for training. This is not surprising due to the explicit focus of the CH construction procedure on excluding poorly predicted systems.

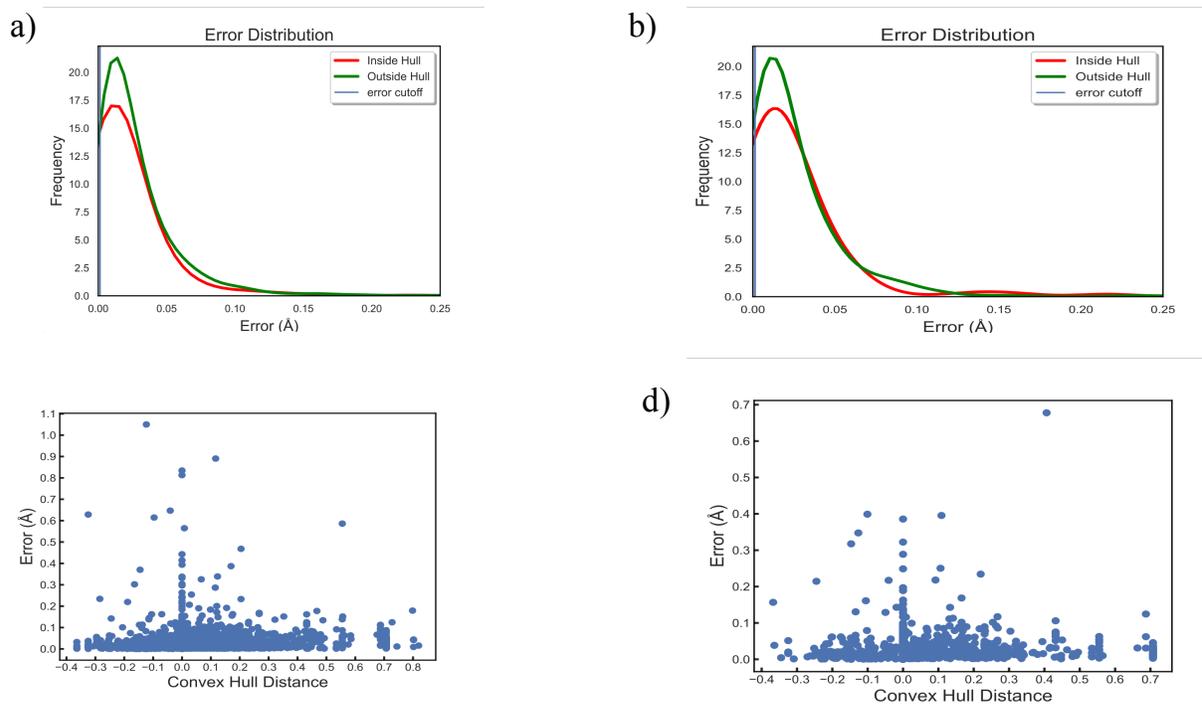

**Figure S1. Results for convex hull construction using the 5% of the data points with the lowest error for the perovskite oxide lattice parameters** a) Error distributions for systems inside and outside the convex hull for the training set used in CH construction  b) Error distributions for systems inside and outside the convex hull for the test set not used in CH construction.   c) Errors plotted versus the distance from the CH boundary for the training set used in CH construction   d) Errors plotted versus the distance from the CH boundary for the test set not used in CH construction

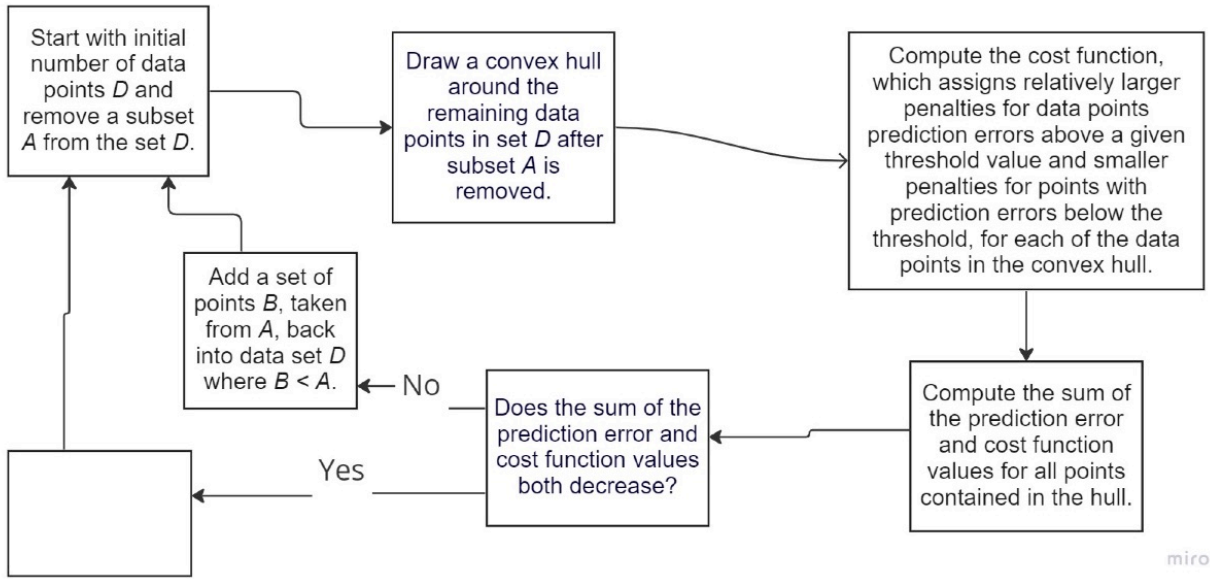

**Figure S2. Flow chart of the optimization procedure used to construct a convex hull that includes the highest possible well-predicted systems and excludes poorly-predicted systems.**

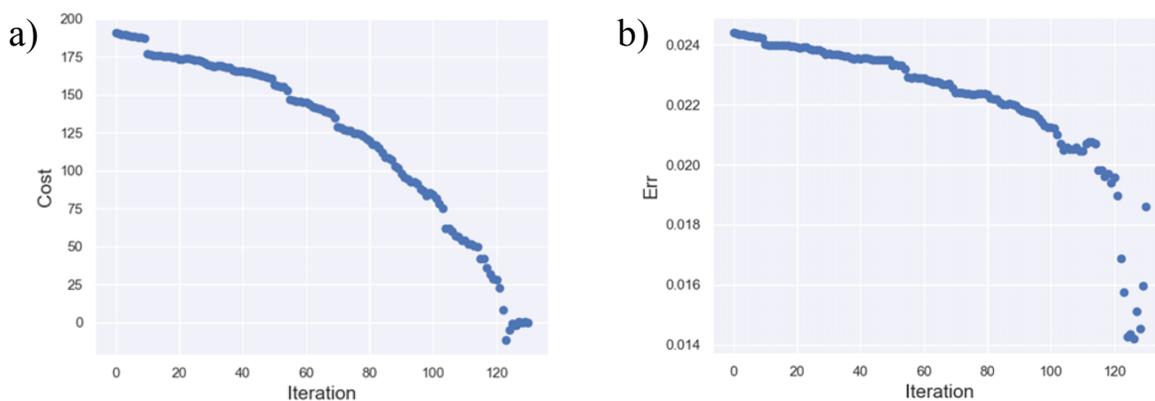

**Figure S3.** a) Cost function as a function of iteration b) MAE of the ABO$_3$ systems chosen by the convex hull construction algorithm as a function of iteration

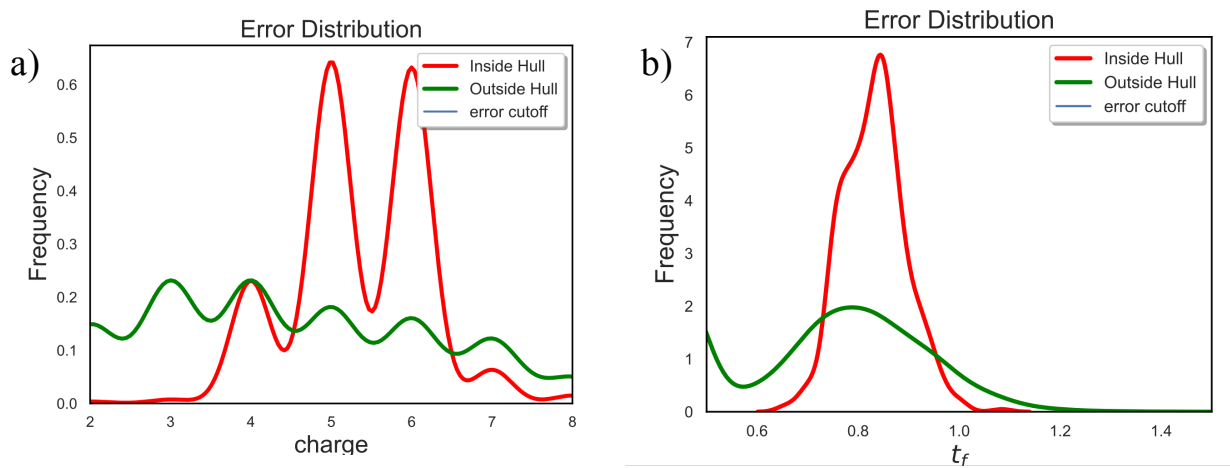

**Figure S4. Distributions of the total cation charge and tolerance factor for the perovskite oxides systems inside and outside the constructed convex hull** a) Distributions of the total cation charge b) Distributions of the tolerance factor

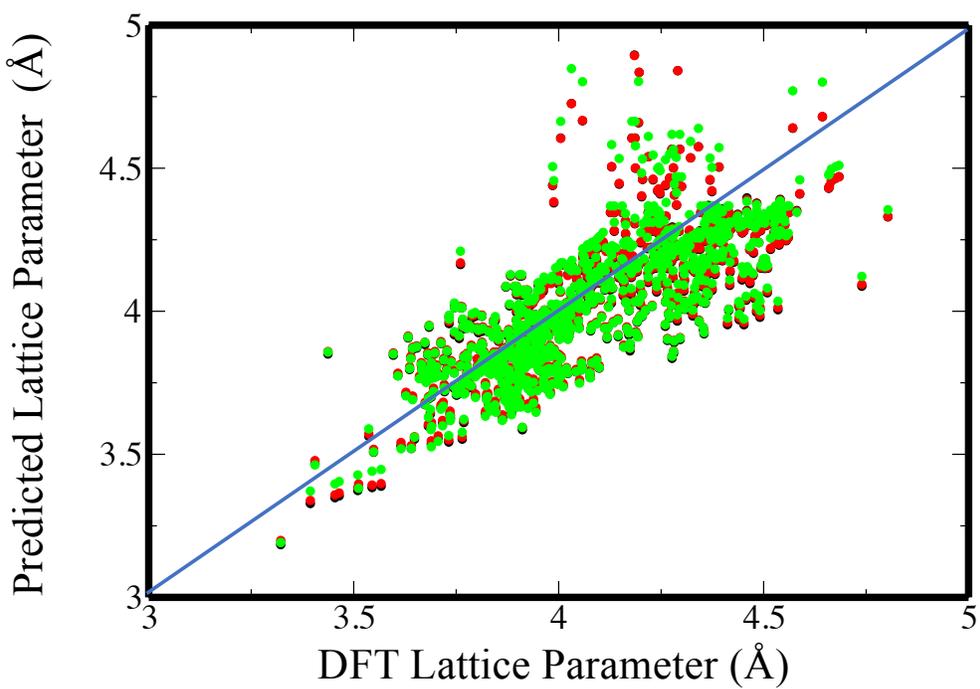

**Figure S5. Predictions of analytical models for perovskite lattice parameters** a) Predictions of Analytical models for ABO₃ lattice parameter compared to the DFT calculated values. The data for the Sidey [20], Ubic [19] and Jiang [21] models shown in black, red and green,

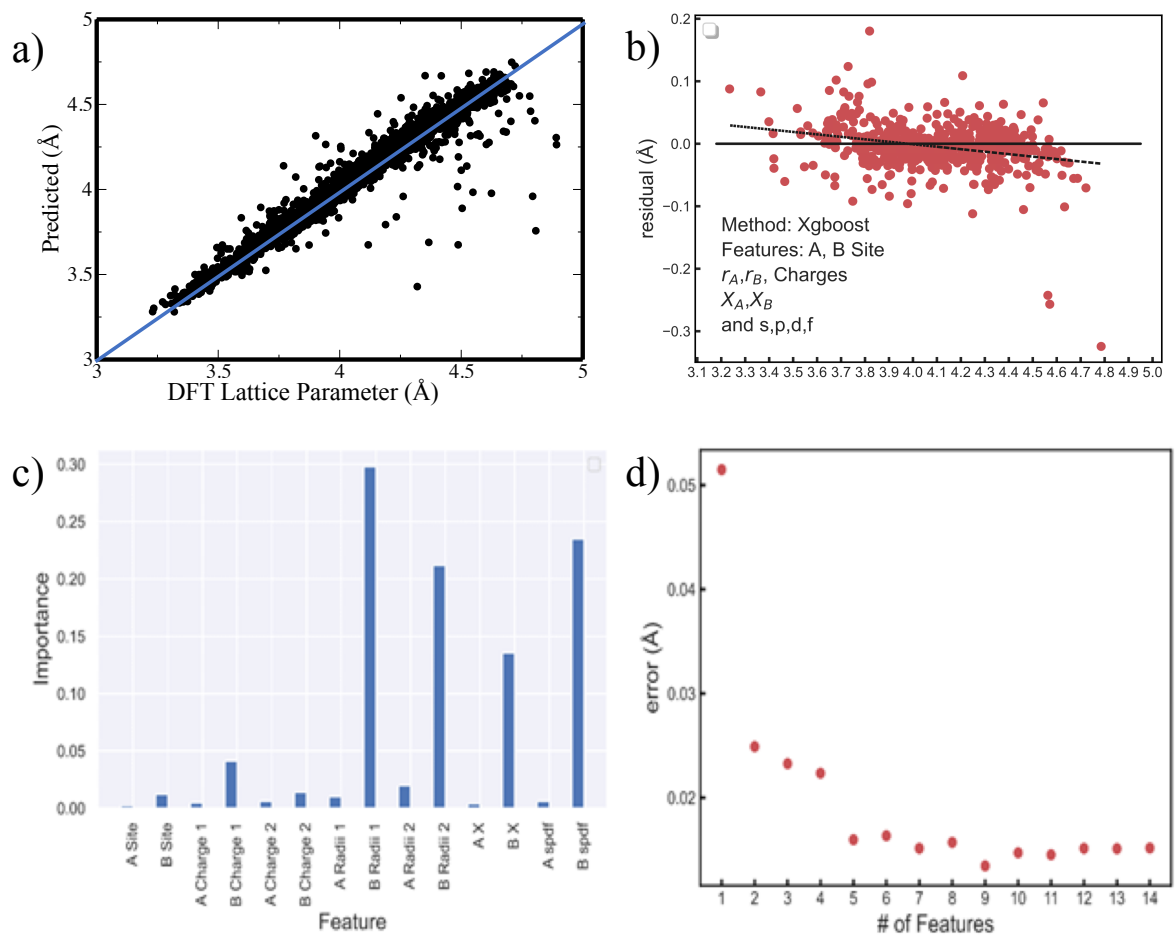

**Figure S6. Characteristics of the XGboost model used for perovskite oxide lattice parameter prediction** a) Predicted versus DFT lattice parameter for our XGboost ML model trained on all data points. b) Residuals for our XGboost model as a function of lattice parameter c) Feature importance plot for our XGboost model d) Error as a function of number of features for our XGboost model

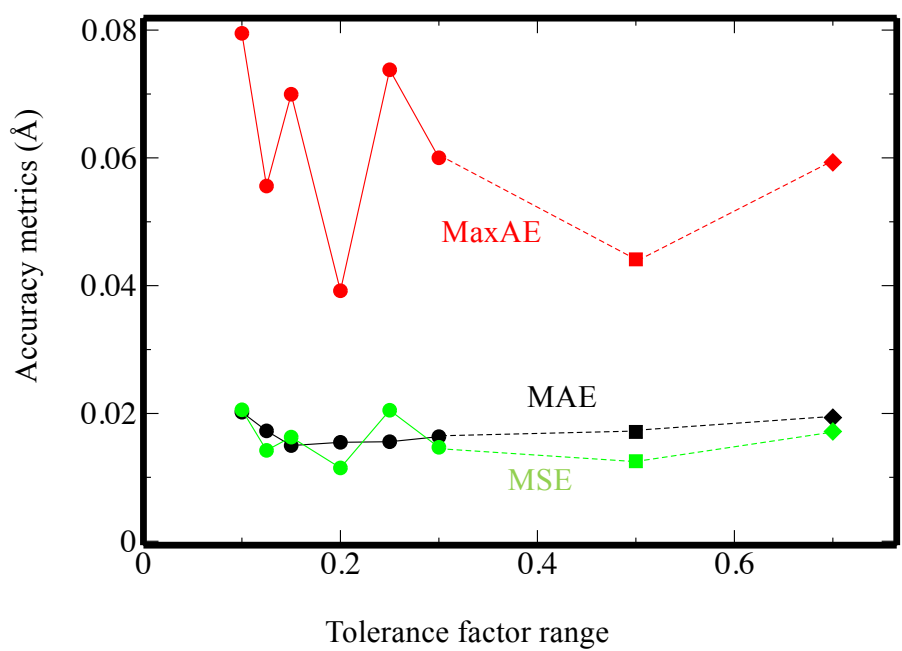

**Figure S7.** a) MAE, MaxAE and MSE for the experimentally-viable systems with 0.9≤t≤1.1 as a function of tolerance factor deviation from 1.0 of the ABO₃ systems used for training of the XGboost model (circles). MAE, MaxAE and MSE for the experimentally-viable systems with 0.9≤t≤1.1using all charge-balanced ABO3 systems for XGboost training (squares). MAE, MaxAE and MSE for the experimentally-viable systems with 0.9≤t≤1.1using all ABO3 systems for XGboost training (diamonds).

**Table S1** Errors for lattice parameter predictions on ABO₃ systems for analytical models and machine learning methods.

| Method | MAE (Å) | MSE (Å²) | stdevError (Å) | MaxAE (Å) |
|---|---|---|---|---|
| Sidey, Analytical, all points | 0.142 | 0.038 | 0.11043 | 0.71078 |
| Jiang, Analytical, all points | 0.146 | 0.044 | 0.11561 | 0.87062 |
| Ubic, Analytical, all points | 0.140 | 0.037 | 0.10948 | 0.70892 |
| Olowabi, SVRA-PSO, all points | 0.029 | 0.00336 | 0.0503 | 1.08432 |
| Fingerprint CNN, all points | 0.037 | 0.00282 | N/A | N/A |
| Xgboost, all points | 0.026 | 0.00289 | 0.04690 | 1.04991 |

**Table S2**: Best Features for each number of optimal features

| # of Features | Best Features |
|---|---|
| 1 | B Site Label |
| 2 | B Site Label and smaller A Radius |
| 3 | B Site Label, smaller A Radius, A X |
| 4 | B Site Label, smaller A Radius, smaller B Radius, A X |
| 5 | B Site Label, larger A Site Charge, smaller A Radius, smaller B Radius, A X |
| 6 | B Site Label, larger A Site Charge, smaller B Radius, smaller A radius, A X, B X |
| 7 | B Site Label, smaller B Site Charge, larger A Site Charge, smaller B Radius value, smaller A Radius , A X, B X |
| 8 | B Site Label, smaller B Site Charge, larger A Site Charge, smaller B Radius , smaller A Radius , A X, B X and A Site s,p,d,f block |
| 9 | B Site Label, smaller B Site Charge, larger A Site Charge, smaller and larger B radius, smaller A Radius, A X, B X and A Site s,p,d,f block |

**Table S3**: ML accuracy metrics for the predicted $ABO_3$ lattice parameter for various data sets for the XGboost model trained on the full set of $ABO_3$ systems.

| Set of Data Points | MAE (Å) | stdError (Å) | MaxAE (Å) |
|---|---|---|---|
| All Oxide Points | 0.026 | 0.0469 | 1.0499 |
| Manually Chosen Experimentally Viable Points | 0.018 | 0.0187 | 0.1135 |
| Convex Hull Defined Points | 0.015 | 0.0179 | 0.1159 |
| Manually Chosen and Convex Hull Defined Points | 0.017 | 0.0184 | 0.1159 |

**Table S4**: ML accuracy metrics for the full, experimentally viable, and convex-hull data set for the XGboost models trained on the full, experimentally viable, and convex-hull data sets of ABO3 systems.

|  | All Oxides Test (Å) | Experimentally viable Test (Å) | Convex Hull Defined Test (Å) | Human and Convex Hull Defined Test (Å) |
|---|---|---|---|---|
| All Oxides Train | 0.026 | 0.018 | 0.015 | 0.017 |
| Experimentally viable Train | 0.11 | 0.021 | **0.088** | 0.062 |
| Convex Hull Defined Train | 0.074 | **0.096** | 0.018 | 0.031 |
| Human and Convex Hull Defined Train | 0.061 | 0.022 | 0.014 | 0.017 |